# Characterization of the Surface Charge Property and Porosity of Track-etched Polymer Membranes

Jiakun Zhuang,[1,2] Long Ma,[1,2] and Yinghua Qiu[1,2,3,4]*

1. Key Laboratory of High Efficiency and Clean Mechanical Manufacture of Ministry of Education, National Demonstration Center for Experimental Mechanical Engineering Education, School of Mechanical Engineering, Shandong University, Jinan, 250061, China
2. Shenzhen Research Institute of Shandong University, Shenzhen, Guangdong, 518000, China
3. Suzhou Research Institute, Shandong University, Suzhou, Jiangsu, 215123, China
4. Key Laboratory of Ocean Energy Utilization and Energy Conservation of Ministry of Education, Dalian, Liaoning, 116024, China

*Corresponding author: yinghua.qiu@sdu.edu.cn

**Abstract**

As an important property of porous membranes, the surface charge property determines many ionic behaviors of nanopores, such as ionic conductance and selectivity. Based on the dependence of electric double layers on bulk concentrations, ionic conductance through nanopores at high and low concentrations is governed by the bulk conductance and surface charge density, respectively. Here, through the investigation of ionic conductance inside track-etched single polyethylene terephthalate (PET) nanopores under various concentrations, the surface charge density of PET membranes is extracted as ~−0.021 C/m$^2$ at pH 10 over measurements with 40 PET nanopores. Simulations show that surface roughness can cause underestimation in surface charge density due to the inhibited electroosmotic flow. Then, the averaged pore size and porosity of track-etched multipore PET membranes are characterized by the developed ionic conductance method. Through coupled theoretical predictions in ionic conductance under high and low concentrations, the averaged pore size and porosity of porous membranes can be obtained simultaneously. Our method provides a simple and precise way to characterize the pore size and porosity of multipore membranes, especially for those with sub-100 nm pores and low porosities.

**Abbreviations**

    PET
        polyethylene terephthalate
    PC

polycarbonate

EDLs
: electrical double layers

PNP
: Poisson−Nernst−Planck

NS
: Navier−Stokes

$R_{total}$
: electrical resistance

$R_P$
: pore resistance

$R_{ac}$
: access resistance

## 1. Introduction

Besides biological nanopores, solid-state nanopores have attracted much attention due to their various advantages in tunable dimensions [1], high stability in solutions, and convenient surface modification [2,3]. These nanopores have promising applications in nanofluidic sensing [4], DNA sequencing [5], energy conversion [6], and desalination [7].

With the track-etching technique, polymer nanopores, such as polyethylene terephthalate (PET) and polycarbonate (PC), can be easily fabricated, which provide an important versatile platform for the investigation of ion and fluid transport in confined spaces [8-11]. In order to reach practical applications, porous polymer membranes are usually applied in osmotic energy conversion [12], and seawater desalination [13]. Taking advantage of the controllable material-etching process, a series of nanopores with different geometries can be achieved, for example cylindrical [9], conical [14], funnel-shaped [15], bullet-shaped [16], and cigar-shaped [17]. Based on the exposed carboxyl groups on the membrane surface and various chemical modifications [3], the surface properties such as charge polarity [11], charge density [18], and wettability [19] of polymer nanopores can be tuned conveniently. These modifications enable nanopores smart responses to ultraviolet light, temperature, and pH changes [20-22].

The surface charge property is determined by the chemical composition of materials, which influences the transport of ions and fluid under confined spaces significantly due to the formation of electrical double layers (EDLs) and induced electroosmotic flow [23]. The amount of surface functional groups can be evaluated quantitatively with the photoinitiator characterization [24] and inductively coupled plasma mass spectrometry [25]. Because nanopores have surface roughness, the surface charge density considered below is the effective value in real cases.

The surface charge density of −0.16 C/m$^2$ was used for PET surfaces because one carboxyl group was thought to be created per square nanometer when the polymer chain was broken during the etching process [26]. By simulating the ionic current through conical nanopores, Cervera et al. [27] obtained the surface charge density of PET membranes as −0.048 C/m$^2$ under pH 5.6. However, their simulation results may not represent the actual situation, because Navier-Stokes equations which describe the induced electroosmotic flow in nanopores were not included [28]. During these characterizations, the surface roughness of PET membranes [10,29] was not taken into consideration which may affect the electroosmotic flow [30] and produce obvious errors in prediction. With the streaming potential method, i.e. measuring the ionic current under a hydrostatic pressure applied across nanopores, Déjardin et al. [31] evaluated the surface charge density as −0.012 C/m$^2$ with multitrack PET membranes in 10 mM KCl solution. While through the same way, Xue et al. [32] obtained the surface charge density varying from ~−1 to −10 mC/m$^2$ for a single PET nanopore, because the zeta potential near charged pore walls is determined by both the surface charge density and the salt concentration. Due to the complicated experimental setup, difficult pressure control, and the unclear relationship between the zeta potential and the surface charge density [23,33], the streaming potential method may not be a best choice to characterize the surface charge properties of nanopore surfaces.

The Debye length ($\lambda$) describes the thickness of EDLs through $\lambda=\sqrt{\dfrac{\varepsilon_0 \varepsilon kT}{2e^2 \rho_\infty}}$ in monovalent electrolyte solutions, where $\varepsilon_0$, $\varepsilon$, $k$, $T$, $e$ and $\rho_\infty$ are the vacuum permittivity, dielectric constant of water, Boltzmann constant, temperature, elementary charge, and bulk concentration, respectively [23,33]. In solutions with a concentration lower than 0.1 mM which induces a Debye screening length larger than 30 nm, EDLs near charged walls dominate the main space or may overlap in nanopores with sub-100 nm diameters. The ionic transport in these cases is determined by surface charges totally, that induces surface-charge-governed ionic current [34-36]. In more diluted solutions, the surface-charge-governed ionic current keeps at a constant value, which can be predicted theoretically from the stable ionic concentrations controlled by the surface charge density [37]. Due to the high accuracy and simple operation procedures in the measurement, the ionic conductance method can provide a convenient way to detect the surface charge density [35,38].

For porous membranes, the bubble point testing [39] is usually used for the measurement of the pore size and porosity. From the theoretical prediction, in order to size the porous membrane with sub-100 nm nanopores, high gas pressure or solutions with low surface tensions like organic solvents should be used which may cause the membrane rupture under ultra-high pressures or induce degradation and swelling of polymer membranes. Imaging with a scanning electron microscopy (SEM) provides another convenient way for the characterization of nanostructures. However, it's still challenging to characterize nanopores with sub-100 nm. Due to the small vision area at large magnification amplitudes, finding and analyzing those nanopores with sub-100 nm in diameter are difficult, especially for the porous membrane with a relatively low porosity such as $10^5$ pores/cm$^2$. Also, the pore size is determined by the artificial selection of the pore boundary, which is usually blurred at high magnification rates. Further, because of the high energy of electron beam used in SEM, the surface of polymer membranes may be damaged by electron bombardment and deformed by released heat. Please note transmission electron microscopy (TEM) can provide a higher resolution, but its application is limited by the sample thickness which is usually less than a few hundreds of nanometers.

From the ionic conductance measurement, the influences of surface charges are negligible in solutions with a concentration higher than 1 M where the Debye length reaches shorter than 0.3 nm. The obtained ionic current is determined mainly by the cross-section area of nanopores, and bulk conductance [40]. Taking advantage of the highly accurate measurement for ionic current, diameters of individual nanopores can be characterized effectively without high-resolution electron beam microscopies [9,41]. While with the ionic conductance method, as far as we know the pore size and porosity of multipore membranes have not been evaluated.

In this work, the surface charge property of PET nanopores is characterized by the surface-charge-governed ionic conductance. With the analysis of more than 40 PET nanopores, the surface charge density of PET membranes was determined as ~−0.021 C/m$^2$. This value was much less than that used in the literature [14,42], and similar to the experimental results [31,32]. Through finite element simulations with COMSOL Multiphysics, the surface roughness that has seldom been considered notably decreased the surface-charge-governed ionic current which induces relatively lower effective surface charge density. Then, with the extracted surface charge density, two types of porous membranes with $10^5$ and $10^7$ pores/cm$^2$ were tested under low- (less than 0.1 mM) and high-concentration (higher than 0.5 M) solutions. Based on the perfect theoretical predictions of current behaviors in both regions, the pore size and porosity have been obtained simultaneously. The predicted pore size from ionic conductance agrees well with that from SEM imaging which is difficult to be conducted especially for membranes with $10^5$ pores/cm$^2$. Our method provides a convenient and low-cost way for the characterization of porous membranes.

## 2. Experimental and Simulation Methods

Single-track and multiple-track PET membranes used in the experiments were fabricated by the track-etching technique [10,29,42]. PET membranes with 30 mm in diameter and 12 μm in thickness were bombarded with single/multiple heavy ions (GSI Helmholtz Center for Heavy Ion Research, Darmstadt, Germany). After heavy ions passed through the PET membrane, latent tracks were left in the membrane. Before etching, both the front and back sides of PET membranes were sensitized under ultraviolet light with a wavelength of 365 nm (UVP UVGL-25, CA, USA) for 1 h. 0.5 M NaOH solutions were then used for the wet etching of membranes at a 70 ℃ water bath (Julabo VIVO B3, Beijing, China). During the etching process, cylindrical nanopores are created along latent tracks whose diameter is linearly proportional to the etching time [42].

Pore diameters were determined from the ionic conductance in 1 M KCl solution through a picoammeter Keithley 6487 (Keithley Instruments, Solon, Ohio, USA). As shown in Figure 1A, porous PET membranes are sandwiched between two reservoirs so that nanopores provide the only passageway for the transport of ions and fluid. In experiments, KCl solutions with concentrations from $10^{-6}$ to 2 M have been considered. The solution pH was buffered at 10 with 10 mM Trizma base. Solutions with lower concentrations were prepared by diluting concentrated KCl solutions with buffered deionized water. All chemicals used in the experiment were purchased from Sigma-Aldrich unless otherwise noted. Deionized water (18.2 MΩ) was purified by Direct-Q 3UV (MilliporeSigma, Burlington, MA, USA). The conductivity of solutions was measured with a conductivity meter (Mettler Toledo FE 38, Shanghai, China). The current-voltage (IV) curves were obtained by averaging two scans from −0.5 to 0.5 V with a pair of homemade Ag/AgCl electrodes by 0.1 V step. (Figure S1) Then the ionic conductance in nanopores can be extracted at different voltages.

For porous membranes with the nominal density of $10^5$ and $10^7$ pores/cm$^2$, after characterization with the method of ionic conductance, other parts of the same membrane were imaged with SEM (JEOL JSM-7800F, Tokyo, Japan and Carl Zeiss Gemini 300, Oberkochen, Germany), with a gold layer of ~15 nm in thickness (JEOL JEC-3000FC or JEOL JEC-1600, Tokyo, Japan). Please note that the nominal pore density of multi-track membranes was obtained at GSI through rough averaging the latent track number over an area with ~2 cm in diameter.

Three-dimensional finite element simulations were conducted using COMSOL Multiphysics coupled with Poisson−Nernst−Planck (PNP) and Navier−Stokes (NS) equations which were used to describe the EDLs structure, ionic transport, and fluid movement through nanopores [14,43]. The pore length and diameter were set to 12 μm and 150 nm. The surface roughness was considered a uniform nanoscale pattern on the inner-pore walls. As shown in Figure 1B, three different rough patterns used in the simulations are circular, rectangular, and triangular with a height ($R_z$) of 10 nm. This value was referred to the AFM detection by Siwy et al. From their experiment, the surface roughness mainly varied from ~20 to ~80 nm for a ~500 nm diameter pore [29]. Considering the more diluted etching solution used in this work which induces relatively smoother surfaces, averaged roughness height of 10 nm was used for the 150 nm diameter nanopore. Uniform surface charge density was added to the inner and exterior pore walls. KCl solutions were used with the diffusion coefficients of K$^+$ and Cl$^-$ ions as 1.96×10$^{-9}$ and 2.03×10$^{-9}$ m$^2$ s$^{-1}$ [44]. Detailed boundary conditions are provided in Table S1 [45,46]. For charged surfaces, the mesh size of 0.2 nm was used to capture the ionic behaviors in EDLs [45,46]. Please note that the NS equation was not solved in the cases without the consideration of electroosmotic flow. For solutions used in simulations, ionic activity has been considered for each concentration (Table S2) [47].

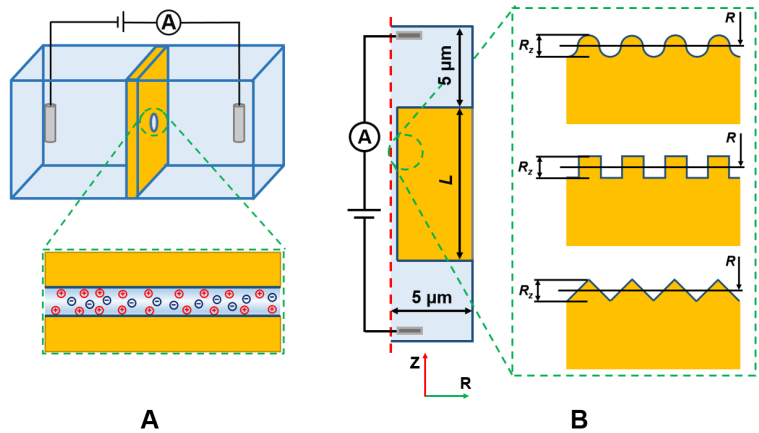

Figure 1 Schematic diagram of the nanofluidic experiment setup and simulation models. (A) Experimental setup with a porous PET membrane sandwiched between two reservoirs. Inset shows the zoomed-in nanopore region. (B) Simulation model for ionic conductance through nanopores. Rough surface patterns for inner pore walls are considered. $L$, $R$, and $R_z$ are the pore length, radius, and roughness height, respectively. Both the length and radius of reservoirs are 5 μm.

## 3. Results and Discussion

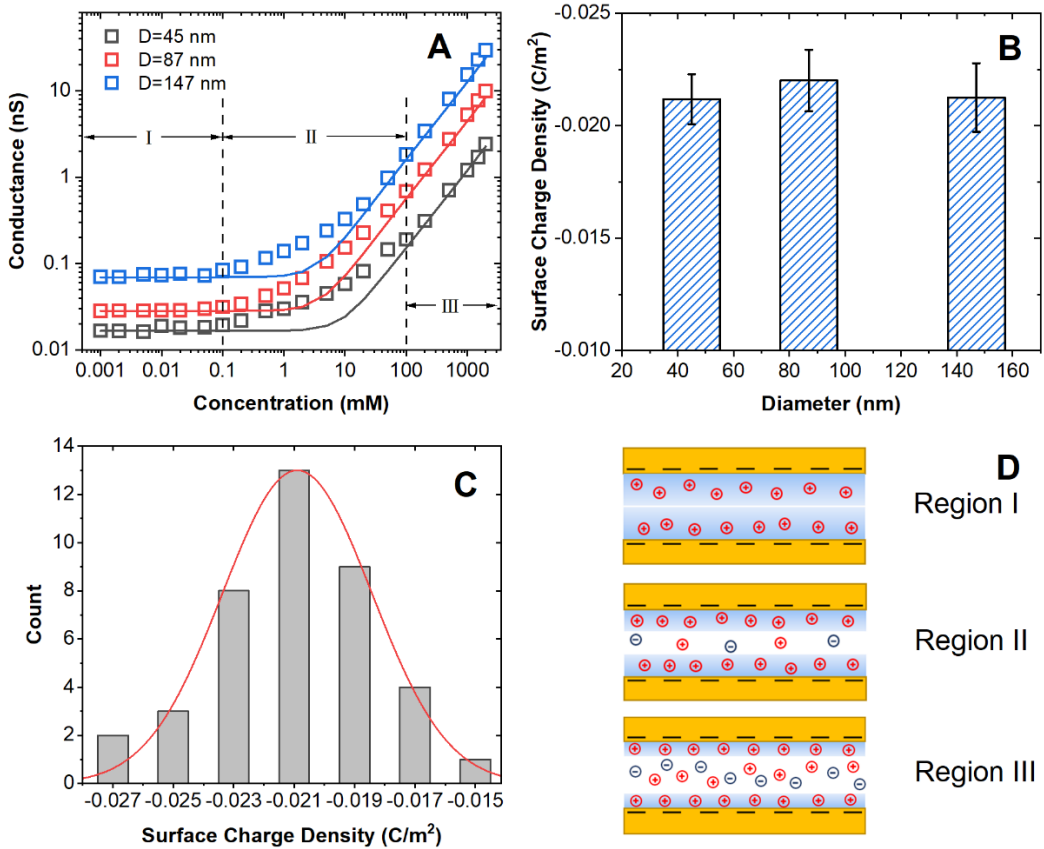

Figure 2 Ionic conductance through single nanopores obtained from nanofluidic experiments and the predicted surface charge density from equations. (A) Ionic conductance under various solution concentrations through three nanopores with different diameters at 0.5 V. Solid lines are theoretical fitting with equation 2. (B) Predicted surface charge density for three nanopores with different diameters. (C) Distribution of measured surface charge density over 40 different PET nanopores. The red line shows the Gaussian fitting. (D) Schematic illustrations of electric double layers inside nanopores under various concentrations. Region I: thick or overlapped EDLs in diluted solutions, Region II: EDLs with a moderate thickness, and Region III: thin EDLs in concentrated solutions. Surface roughness is not shown.

Through nanofluidic experiments with single cylindrical nanopores, ionic current behaviors have been investigated in KCl solutions varying from $10^{-6}$ to 2 M at pH 10.

There is no specific consideration for the application of pH 10. Other pH values work in the same way. In Figure 2A, the ionic conductance profiles share a similar trend through three individual nanopores with different diameters. The conductance can be divided into three regions: [35,36] (I) the surface-charge-governed ionic conductance, (II) the transition region, and (III) the bulk-concentration-determined conductance.

For the KCl concentration larger than ~0.1 M, the ionic conductance increases linearly with the solution concentration (Region III), which is not affected by the surface charge density of nanopores because of the strong electrostatic screening of counterions to surface charges (Figure 2D). The ionic current through nanopores is mainly determined by the solution conductivity and pore diameter. For cylindrical nanopores, the total electrical resistance ($R_{total}$) includes two parts: the pore resistance ($R_P$), and the access resistance ($R_{ac}$) as described in equation 1 [48]. Based on the measured current values, the diameter of nanopores can be obtained. Please note that for the nanopores with 12 μm in length, the access resistance takes a little percentage of the total resistance [48].

$$R_{total} = R_P + R_{ac} = \frac{L}{\kappa \pi R^2} + \frac{1}{2\kappa R} \qquad (1)$$

where $L$, $R$, and $\kappa$ are the membrane thickness, pore radius, and solution conductivity, respectively.

In KCl solutions with a concentration lower than ~0.1 mM, the ionic conductance through nanopores keeps at a constant value (Region I) which does not vary with the bulk concentration. The conductance is called surface-charge-governed ionic conductance [35]. This phenomenon is mainly due to the unchanged ionic concentration inside the nanopore which is resulted from the constant surface charge density. In diluted solutions, because of the large Debye lengths the confined space of a nanopore is dominated by EDLs regions near charged walls that extend to the pore center or even overlap inside the nanopore (Figure 2D). In these cases, the nanopore mainly accommodates counterions, and the intra-pore ionic concentration is determined by the surface charges based on electrical neutrality [23].

Considering the modulation effect from the surface charge density on the concentration of counterions inside nanopores in equation 1, the ionic conductance ($G$) through charged nanopores can be described with the mean-field equation 2 [35,37,49].

$$G = FC\gamma(\mu_+ + \mu_-)\sqrt{1+\left(\frac{\sigma}{FC\gamma R}\right)^2}\left(\frac{L}{\pi R^2}+\frac{1}{2R}\right)^{-1} \qquad (2)$$

in which $\mu_+$ and $\mu_-$ are the ionic mobility of K$^+$ and Cl$^-$ ions, equal to 7.616×10$^{-8}$ and 7.909×10$^{-8}$ m$^2$s$^{-1}$V$^{-1}$, respectively [50]. $\sigma$, $F$, $C$, and $\gamma$ are the surface charge density, Faraday's constant, bulk concentration, and activity coefficient, respectively. Please note that for concentrated solutions, activity coefficients should be considered due to the enhanced ionic interaction between cations and anions which decreases the actual conductivity of solutions [47]. Table S2 lists the activity coefficients for the solutions used in the experiments and following simulations [50]. From Figure S2, for the surface-charge-governed conductance, the contributions caused by surface charges to the ionic conductance can reach ~100% in solutions with a concentration lower than 1 mM in a 147 nm diameter nanopore.

As shown in Figure 2A, the ionic conductance located in regions I and III can be predicted precisely based on the good agreement between the experimental data and theoretical results with equation 2. For nanopores created on PET foils with 12 μm in thickness, with the predicted pore radius from the ionic conductance in region III, the surface charge density can be evaluated through the data in region I using equation 2. Because the surface charges on PET surfaces are induced by the protonation/deprotonation of carboxyl groups, the surface charge density of pore walls keeps at the same value in solutions with the same pH but different salt concentrations [33,35,36]. From Figure 2B, for three PET pores with different diameters, the surface charge density was evaluated which does not exhibit clear dependence on the pore size varying from ~50 to ~150 nm. Following the same strategy, the surface charge density has been measured with 40 nanopores to provide an accurate investigation of the surface charge property of PET nanopores. As shown in Figure 2C, the obtained surface charge density of PET membranes follows a Gaussian distribution with the center at ~−0.021 C/m$^2$.

The experimental conductance through nanopores obtained in region II is slightly higher than the theoretical prediction. With a solution concentration located in this region, EDLs near charged surfaces have a moderate thickness (Figure 2D). In these cases, both the surface-charge-governed conductance and bulk-concentration-determined conductance can have a considerable contribution to the total ionic conductance, and it's complicated to identify the percentage of each contribution [23,35]. Because the ionic conductance in regions I and III is important to the evaluation of the surface charge density and pore size, here we haven't attempted to provide a detailed investigation of the ionic conductance in region II.

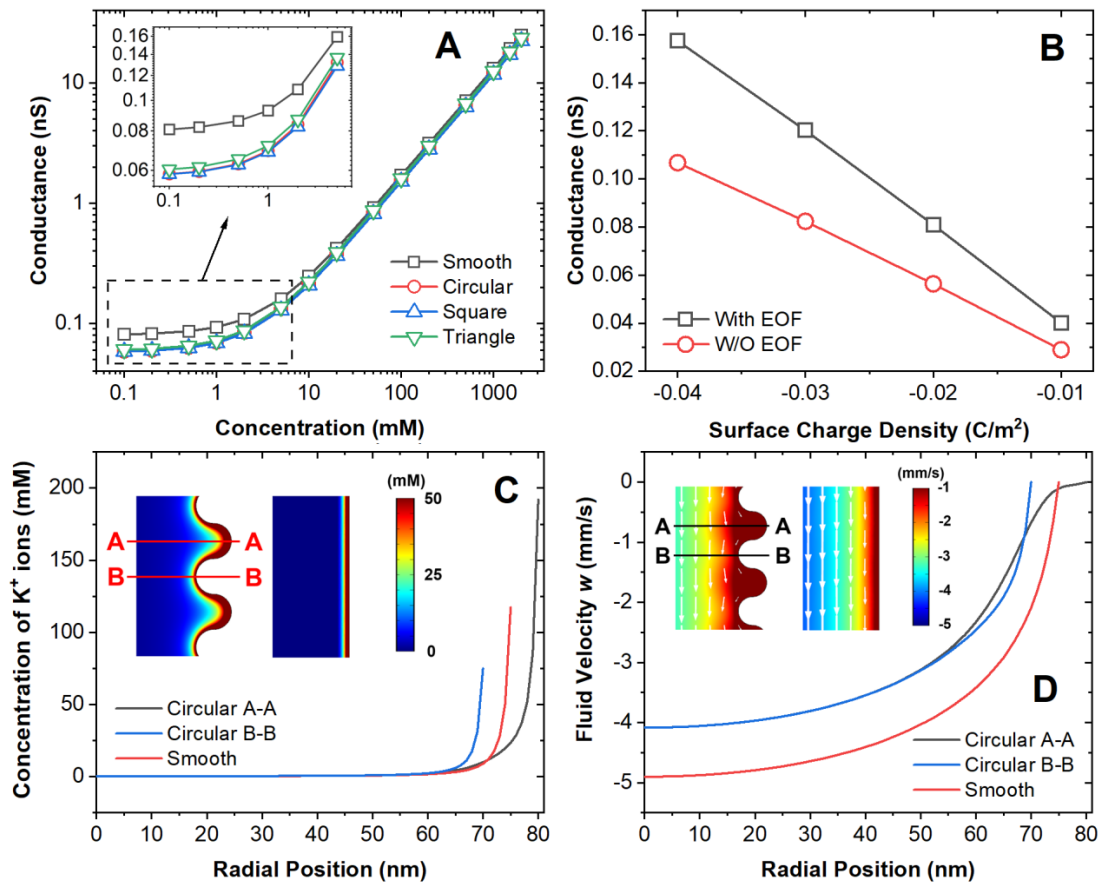

Figure 3 Simulated ionic conductance, as well as transport of ions and fluid through nanopores with 150 nm in diameter and 12 μm in length. (A) Ionic conductance under various bulk concentrations from simulation models with smooth and rough inner-pore walls. Surface roughness patterns are shown in Figure 1b. (B) Ionic conductance under different surface charge densities from simulations with and without electroosmotic flow (EOF). (C-D) Distributions of the ion concentration (C) and fluid velocity (D) in the radial direction obtained from the center cross-sections of the nanopores with rough and smooth surfaces. The white arrows in (D) show the fluid flow direction, and their size denotes the flow speed. The solution was 0.1 mM KCl. The surface charge density and voltage were set to −0.02 C/m² and 0.5 V, respectively.

The evaluated surface charge density of PET membranes at pH 10 is ~−0.021 C/m²

with the ionic conductance method. While we think our method may be influenced by the surface roughness because surface roughness can affect the ionic conductance at region I by enlarging the surface area and slowing down the electroosmotic flow (EOF) inside the nanopore, simultaneously. Please note that surface roughness is the inherent property of the membrane surface which is difficult to be considered in theoretical predictions. In all earlier experimental work with the stream potential method [31,32], and this work, the obtained surface charge densities are all effective values, which are based on the assumption that cylindrical nanopores have smooth inner-pore walls in the theoretical analysis.

Here, referring to the roughness of PET surfaces detected by atomic force microscopy [29], different simulation models have been built with rough inner-pore walls (Figure 1B). The influences from surface roughness on the ionic conductance were investigated through nanofluidic simulations in KCl solutions varying from 0.1 to 2000 mM in concentration. As shown in Figure 3A, the surface-charge-governed ionic conductance appears at 0.2 mM. In the 150 nm in diameter nanopores, surface roughness can inhibit the surface-charge-governed ionic conductance by ~25%, though it expands the charged surface area inside the nanopore which can attract more counterions to increase the ion concentration (see below). In region I, the suppression from surface roughness in the ionic conductance does not show a clear dependence on the roughness pattern. While at region III surface roughness has no obvious effects on the ionic conductance.

In the simulations, with the consideration of fluid movement, the EOF induced by the directional transport of cations can be investigated. From Figure 3B, the contribution of EOF to the ionic conductance at region I is evaluated through simulation cases with and without NS equations. We can see that EOF plays an important contribution to ionic conductance. With the surface charge density varying from −0.01 to −0.04 C/m$^2$, EOF contributes ~30% of the total ionic conductance. As shown in Figure S3, the promotion of EOF to the ionic current has been investigated with nanopores of different diameters. With the pore size increasing from 50 to 500 nm, the enhancement from EOF to the surface-charge-governed ionic conductance reaches above 35%.

EOF originates from the directional movement of counterions in EDLs with an applied voltage across the nanopore due to the hydration effect [51], which in turn promotes the movement of counterions [28,52]. Under the non-slippery boundary conditions, surface roughness could have a hindrance on fluid flow by decreasing the ionic transport and increasing the flow resistance [30,53,54]. The inhibited EOF is responsible for the lower ionic conductance in rough nanopores. Detailed influences from the surface roughness on ionic transport are shown in Figures 3C and 3D. For the rough nanopore with the circular pattern, from the inset of Figure 3C, the EDLs exhibit non-uniformity near the valley and summit regions on the rough pore walls. Compared to the case with smooth pore surfaces, much more counterions accumulate in the valleys of the rough surface. However, most of these counterions do not participate in the ionic transport due to constraints from the surface roughness, which has a negligible contribution to the EOF (Figure 3D) or the total conductance. Because of the protruding topography near the summit regions, the concentration of K$^+$ ions is relatively lower due to the weaker electrostatic attraction between surface charges and free ions, as well as the faster ionic transport in the stronger electric field under the narrower space. For the distribution of axial flow speed perpendicular to the summit regions, though the EOF inside EDLs shares a similar speed to that near smooth walls, it is much slower in the center of the rough nanopore. Then, the smaller surface-charge-governed ionic conductance through rough nanopores is obtained, which results in a relatively lower surface charge density. Through extra simulations with nanopores of the same diameter but different roughness heights, more rough surfaces can cause larger inhibition to the EOF and surface-charge-governed ionic conductance (Figure S4).

Porous membranes with millions of nanopores are usually required for practical applications in the fields of osmotic energy conversion [55], and seawater desalination [7]. To characterize the properties, such as pore size and porosity, of porous membranes, we developed a method based on ionic conductance obtained in regions I and III in Figure 2A. In solutions with a concentration over 100 mM, surface charges are screened well by counterions, and the total conductance through porous membranes can be treated as the sum of conductance from individual nanopores. While, in the cases with ultra-low salt concentrations, for nanopores with several micrometers in length and sub-200 nm in diameter, the ionic conductance is determined only by the surface charge density of inner-pore walls [35,36]. For a membrane with porosity lower than ~$10^{10}$ pores/cm$^2$, i.e. the averaged pore-to-pore distance larger than ~200 nm, the total conductance can also be treated as the integration over all individual nanopores. Please note that low voltages should be applied to avoid unnecessary concertation polarization across the membrane which can affect the ionic conductance through each nanopore [46,56].

Following the above strategy, referring to equation 2, the ionic conductance through a porous membrane at regions III and I can be theoretically described with equations 3 and 4.

$$G_H = NG_{HS} = NFC_H\gamma_H(\mu_+ + \mu_-)\sqrt{1^2 + (\frac{\sigma}{FC_H\gamma_H R})^2}(\frac{L}{\pi R^2} + \frac{1}{2R})^{-1} \quad (3)$$

$$G_L = NG_{LS} = NFC_L\gamma_L(\mu_+ + \mu_-)\sqrt{1^2 + (\frac{\sigma}{FC_L\gamma_L R})^2}(\frac{L}{\pi R^2} + \frac{1}{2R})^{-1} \quad (4)$$

in which the subscripts H and L represent the values under high and low concentrations located in regions III and I, respectively. $G_H$, and $G_L$ are the total ionic conductance through porous membranes. $G_{HS}$ and $G_{LS}$ denote the corresponding conductance through individual nanopores on porous membranes. $N$ is the pore amount on the exposed membrane area during the measurement of ionic conductance.

$$\frac{G_H}{G_L} = \frac{NG_{HS}}{NG_{LS}} = \frac{C_H \gamma_H \sqrt{1^2 + (\frac{\sigma}{FC_H \gamma_H R})^2}}{C_L \gamma_L \sqrt{1^2 + (\frac{\sigma}{FC_L \gamma_L R})^2}} \quad (5)$$

With coupled equations 3 and 4, equation 5 can be achieved in which only $R$ is the unknown parameter. Because the surface charge density is the inherent property of materials, the surface charge density on PET membranes with a single nanopore and multiple nanopores are believed the same. Using the extracted surface charge density from single nanopores −0.021 C/m², with equation 5 the averaged pore radius can be obtained. Then, the number of nanopores will be found by dividing the total ionic conductance by the conductance in individual nanopores predicted from equation 2.

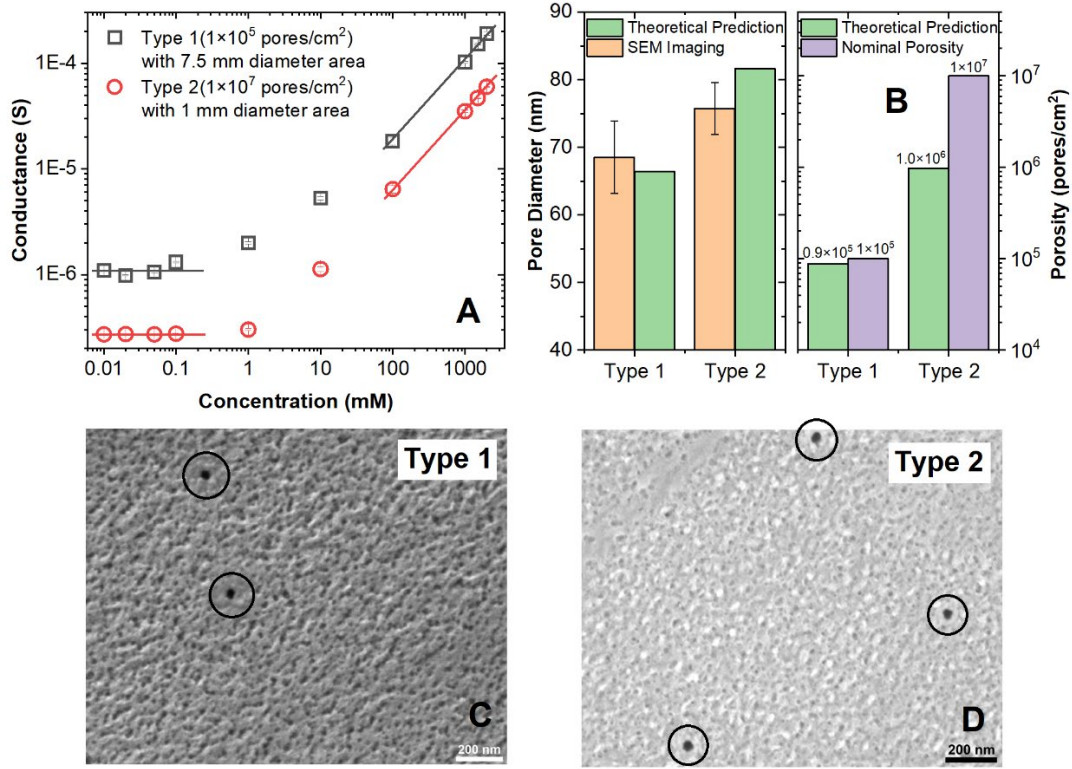

Figure 4 Characterization of porous PET membranes with the ionic conductance method and SEM imaging. Two kinds of porous PET membranes have been tested with the nominal porosity of ~$10^5$ pores/cm$^2$ (Type 1) and ~$10^7$ pores/cm$^2$ (Type 2). (A) Ionic conductance under various KCl concentrations. Exposed membrane sizes are 7.5 and 1 mm in diameter during the conductance measurement with two types of porous membranes. (B) Predicted pore size and porosity with equations 3-5. (C-D) SEM images of both porous membranes. Scale bars in SEM images are 200 nm. Deposited gold layers on porous membranes are ~15 nm for SEM imaging. The nominal density was obtained by averaging the number of tracks over an area with ~2 cm in diameter.

With the developed method, two kinds of porous PET membranes have been characterized with a nominal density of ~$10^5$ and ~$10^7$ pores/cm$^2$ which are denoted in Figure 4 as type 1 and type 2, respectively. Please note that the nominal pore density was a very rough value, which was obtained by averaging the number of latent tracks over a circular area with ~2 cm in diameter during the bombardment by heavy ions at GSI. First, we conducted the measurement of ionic conductance through porous membranes under various salt concentrations from 0.01 to 2000 mM. In the experimental setup, the exposed membrane areas to the aqueous solutions were 7.5 and 1 mm in diameter for porous membranes with a nominal pore density of ~$10^5$ and ~$10^7$ pores/$cm^2$, respectively. The smaller exposed membrane area for porous membranes with denser nanopores was used to induce a smaller electrochemical current at electrodes. As plotted in Figure 4A, the ionic conductance through both porous membranes shares the same trend as that in single nanopores. Then, with the derived equations 3-5, for the membrane of ~$10^5$ pores/$cm^2$, the predicted pore diameter and porosity are ~66.4 nm and ~$0.9 \times 10^5$ pores/$cm^2$. Please note that the obtained porosity is a local density of nanopores which may be different in various regions due to the uneven track distribution on the membranes. While due to the stable etching speed of PET materials [9], nanopores on porous membranes share a similar diameter. After the nanofluidic measurement, the porous membrane was imaged with an SEM. As shown in Figures 4B and 4C, the averaged pore size is ~68.5 nm (Table S3) which agrees well with the predicted value. We need to point out the difficulty during the SEM imaging of the porous membrane with ~$10^5$ pores/$cm^2$. Because of the low porosity, it's very hard to find many nanopores on the membrane. Also, due to the small pore size and blurred pore boundaries, obtaining an accurate pore size becomes a challenging task during the analysis of SEM images.

For a PET membrane with a larger porosity as ~$10^7$ pores/$cm^2$, from the obtained ionic conductance (Figure 4A) the averaged pore size is predicted as ~81.6 nm, which also agrees well with that from SEM characterization ~75.7 nm (Table S3). The obtained local porosity is ~$1.0 \times 10^6$ pores/$cm^2$. This value is close to the nominal pore density. During the conductance measurement, because the exposed membrane area is 1 mm in diameter which takes only 0.25 percent of the total membrane area, we think the potential uneven latent tracks on the porous membranes may induce a larger or smaller local porosity. To confirm our guess, another region was characterized on the same PET membrane (Figure S5). The obtained pore diameter is 75.8 nm, but the local porosity is only ~$2.2\times10^5$ pores/cm$^2$ which is around two orders less than the nominal value. We can find that for the track-etched porous membranes, the porosity may have a significant variation in different local regions.

With equations 3-5, a precise and convenient method has been developed here to characterize the pore size and porosity of porous membranes, simultaneously. The only required input is two measurements of ionic conductance in a high- (such as 1 M) and low-concertation (such as 0.05 mM) solution. Our method may find wide applications in the accurate characterization of porous membranes with small even nanopores and lower porosities. For track-etched membranes with high porosities, the appearance of nanopore overlapping or large uneven nanopores can induce inaccurate measurement of the pore size and porosity [38].

## 4. Conclusions

Due to the electrostatic interactions between surface charges and free ions, EDLs can form near charged pore walls whose thickness is determined by the salt concentration. For ultra-high and ultra-low salt concentrations, thin and thick EDLs represent good and poor electrostatic screening from counterions to surface charges. In both cases, the corresponding ionic conductance through nanopores is determined by the bulk concentration and surface charge density, respectively. Through the investigation of the ionic transport in 40 PET nanopores under various bulk concentrations from $10^{-6}$ to 2 M, the surface charge density of PET membranes, considering nanopores as perfectly cylindrical pores, has been extracted as ~−0.021 C/m$^2$, which does not change with pore sizes. Simulation results show that surface roughness can cause an underestimation of the surface charge density, due to the suppressed EOF which induces smaller surface-charge-governed ionic conductance. With the predicted surface charge density, we developed a method to characterize the averaged pore size and porosity of multipore PET membranes based on the ionic conductance under high and low salt concentrations. For two kinds of porous membranes with a nominal pore density of $10^5$ and $10^7$ pores/cm$^2$, the predicted average diameters are ~66.4 and ~81.6 nm, which are in good agreement with those obtained from SEM imaging. The corresponding local porosities of detected areas are predicted as ~$0.9 \times 10^5$ and $1.0 \times 10^6$ pores/cm$^2$, respectively, which may be affected significantly by the uneven distribution of latent tracks on the membrane. With our methods, the pore size and porosity can be conveniently and precisely obtained, especially for the porous membranes with sub-100 nm nanopores and a low porosity which are difficult to be characterized by SEM imaging.

**Conflicts of interest**

There are no conflicts to declare.

**Acknowledgments**

This research was supported by the National Natural Science Foundation of China (52105579), the Natural Science Foundation of Shandong Province (ZR2020QE188), the Basic and Applied Basic Research Foundation of Guangdong Province (2019A1515110478), the Natural Science Foundation of Jiangsu Province (BK20200234), the Open Foundation of Key Laboratory of Ocean Energy Utilization and Energy Conservation of Ministry of Education (Grant No. LOEC-202109), the Qilu Talented Young Scholar Program of Shandong University, and Key Laboratory of High-efficiency and Clean Mechanical Manufacture at Shandong University, Ministry of Education. We thank the core facilities sharing platform at Shandong University for SEM imaging. The results presented here are based on a UMAT irradiation experiment, which was performed at the beamline X0 at the GSI Helmholtzzentrum für Schwerionenforschung, Darmstadt, (Germany) in the frame of FAIR-Phase 0.

**Supporting Information**

Current-voltage curves under various concentrations through a PET nanopore, contribution caused by surface charges to the ionic conductance, experimental ionic conductance through a multipore membrane, boundary conditions in simulation, activity coefficients and activities of KCl solutions, and pore sizes evaluated from SEM imaging.

**Data Availability Statement**

The data that support the findings of this study are available from the corresponding author upon reasonable request.

**Reference：**

## TOC Graphic

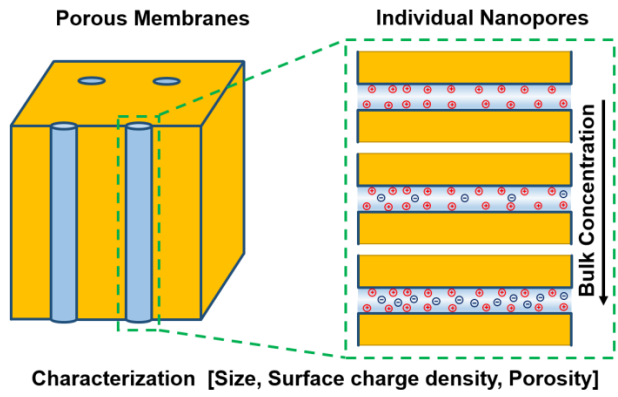